# The role of immersive informal science programs

Jacob Noel-Storr (Columbia University & Science Camp Watonka)
jake@astro.columbia.edu

**Abstract**

Immersive informal environments (such as summer camps, residential programs at museums and science centers, etc.) can provide a venue for young people to explore their scientific thinking in a less formalized context than is available in most traditional classrooms. While class instruction is beneficial for children to develop formal science skills and content knowledge, venues that offer more opportunities for experimentation and exploration can promote deeper understandings. In this article I explore the background of science learning and venues where this learning can take place followed by a review of the benefits and necessary components of well designed immersive informal programs.

## 1. Introduction

A decade ago, Project 2061 of the American Association for the Advancement of Science strongly recommended in their Benchmarks for Science Literacy (AAAS, 1993) that everyone be enabled, through their education, to develop a scientific world view, learn and use the principles of scientific inquiry and understand the scientific enterprise. Working towards achieving the goals presented in that document has proven a significant challenge – prompting a great deal of reform in the way we picture and conduct science instruction. One thing that we have learnt is that in order to turn our children into scientifically literate members of today's modern communities we must offer a complete package of educational experiences and opportunities in a range of settings.

Science learning can take place in all manner of venues. Just the simple day to day experience of the world around us causes us to construct new connections in our minds based on prior understandings that allow us to understand more of the world around us (see e.g. Mestre & Cocking, 2000). This is informal education at its most extreme, where no program has been designed, and there are no goals and outcomes... on the other hand, this learning is a completely personal experience, and it is when we have a personal interest that we are most likely to learn. Even so, how effective a young person's learning is "out in the world" depends strongly on how secure of a scientific foundation we have built for them.

Immersive informal environments are places where young people spend periods of time outside of the classroom immersed in scientific experimentation, investigation and thought. For example many schools now regularly take their students on residential visits where they spend a week or more engaged in learning outside of the classroom; museums and science centers offer programs of substantial duration during school vacations; colleges and universities offer summer programs to high school students; and a collection of summer camps offer specialized programs of this type.

These types of programs, where well designed, can provide experiences in learning that are not accessible in the traditional classroom and can focus on aspects of science education that may be ignored or undervalued by traditional curricula. In this manner they can provide a means of better balancing, rounding and completing a child's science education.

In this article I will first review some aspects of science learning, and discuss some venues where this learning can take place. I will then describe how immersive informal programs can provide a broader range of scientific experiences, followed by a discussion of how good immersive informal programs can be designed and structured.

## 2. Science Learning

The most effective science learning takes place when instruction is designed in an *inquiry based* manner (NRC, 1995; AAAS, 1993). In fact, the cover of the National Science Education Standards stresses the importance of this educational style – dotted with the key words "inquire – observe – learn – reflect – communicate – explore – assess – encourage – interact – understand", without doubt valuable targets for a complete science education. Where this instruction can be focused on the interests of the individual student, the effectiveness of the learning increases still further.

Science learning can be compartmentalized into three areas: knowledge, skills and attitudes. The knowledge component is what many of us gain from our time in school or college. However, serious concern exists about how much of that knowledge remains accessible and useful when taught in a traditional manner. (How many things can you write down that you learnt in $10^{th}$ grade chemistry?)

The transfer of scientific knowledge is a very complex issue in science education. The evolution of scientific ideas strongly suggests that the ability to go through life absorbing and understanding new scientific information is just (if not more) as important as learning particular facts. If we take DNA for example, we would expect every child to come out of school knowing what DNA is, and maybe some other basic information about its structure, its presence in cells, etc... Yet only fifty years ago this would not have appeared on any science curriculum, as its nature had only just been discovered. We would hope though, that scientifically literate adults educated before then could nonetheless appreciate and gain some understanding of that advance.

How scientists understand a particular topic can also often be very different to how students come to understand it. Knowledge is not easily transferred, and each individual must be motivated to come to an understanding of a particular subject in their own mind, making use of the knowledge with which they are provided (Mestre & Cocking, 2002).

There is a strong distinction between knowledge, which a student can be taught, and understanding which they must discover independently (though not necessarily without guidance).

It is now well appreciated that the scientific foundations that an individual needs do not just comprise a set of factual knowledge to be retained throughout our lives. Many educators believe that the skills and attitudes towards science are equally, if not more, important. Content, or the knowledge part of science, only comprises about a sixth of the National Science Education Standards (NRC, 1995). Other areas, such as skills and attitudes, the way science is taught and understanding is assessed, and how science programs are designed and developed, receive equal attention in the standards and are considered equally, if not more, important in developing scientific literacy.

I choose to group scientific skills into two broad sets: the *formal skills* that are expected of a traditional scientist (e.g. using a microscope, performing of a titration, or even plotting a graph); and *informal skills* which are more important in generating understanding from knowledge (e.g. identifying relevant information to use, spotting trends in data and the first stages of developing hypotheses). Informal skills are generally not taught, but young people are expected to pick them up from their experiences. Ensuring that children have access to a sufficient range of experiences to develop these skills is a vital objective.

A good "scientific attitude" is what can eventually enable individuals to think and act scientifically (that is, to generate personal understanding from new knowledge). We are not really interested, in this discussion, in the "attitude towards science" – which can be interpreted in many ways – but rather towards mental attitudes geared towards conducting good science. At the basic level: Are you interested in things and how they interact? This is of course at the core of what it means to have an inquiring mind, and we must make sure that we concentrate on encouraging this kind of attitude.

Inquiry is, by its very nature, a personal process, unique to each individual. If we value inquiry then we need to encourage students to develop the attitudes described above, supported by a basis of knowledge and skills. This type of attitude, along with many of the commensurate skills, is most effectively developed in practical environments, with lively discourse between the participants and instructors (Reardon, 2002). Note that practical environments are not the best place to gain new factual knowledge; but to do anything with the knowledge requires some mental or physical activity.

Individual portions of practical science instruction can be classified as, experiences, exercises or investigations (Woolnough & Allsop, 1985; Noel-Storr, *et al.* 2002). In *experiences*, students get a feeling for equipment or phenomena by manipulating and exploring; in *exercises*, students develop practical skills and techniques by following instructions and recording results; and in *investigations* students act as problem solving scientists.

Such *experiences* in science, resembling a free association between a child and what they are experimenting with, are unfortunately often seen as unstructured activities that have little measurable value. However, time taken to experience science can provide very strong support for knowledge and formal skills by increasing their relevance, and can provide opportunities for an inquisitive attitude to develop (Layman, 1996, chap. 1), more than with much more highly structured exercises and without some of the learning-fear that can be exhibited by children conducting formal investigations.

### 3. Venues for learning

The venue where we learn falls on a spectrum ranging from the most informal (out in the world) – where we rely almost entirely on our previous experiences to construct new scientific understanding – to, traditionally at least, the most formal in schools and colleges. In between lie other experiences, such as visits to museums and science centers, summer camps, and workshops or short courses. The formality of these experiences depends on the particular programs that are offered.

Schools and colleges provide educational environments which can be very well suited to the development of a core base of science knowledge and to the acquisition of formal practical skills (see previous section). Clear goals are laid out by individual instructors, school districts, and at the state and federal levels; little room exists for individuals to pursue their questions and interests.

Without opportunities to learn science outside of formal environments, students may well develop the attitude that science is centered on factual knowledge and formal skills. Most professional scientists would identify themselves as being more inventive and relying more on their inquisitiveness and informal skills while pushing (or even while trying to understand) the frontiers of knowledge.

A visit to a science center or museum provides an environment where "informal" or "free-choice" learning takes place. This learning is very directed by each individual; many may pass through a particular exhibit and appreciate very little of the science, while others make a concerted effort to understand and may learn a lot (Falk, 2001). Informal science education can and should complement the science learning that takes place in schools (Bybee, 2001). In this way students can gain the most from both experiences, providing the programs are designed with this complementary scenario in mind. This, of course, sets up a positive feedback cycle – back in the classroom these students will be more receptive to new knowledge and skills and more able to continue to construct understandings from them.

### 4. Immersive informal programs

I have discussed up to this point the value of learning science in a practical, inquiry driven manner, and the merit of experiencing science education in multiple venues. While science centers and museums provide outstanding opportunities to see a broad range of science in a greater context there is often little space for an individual to truly experiment and to follow their own ideas. For this, a well designed immersive environment can make it possible for children to develop their exploration skills and scientific attitudes in a much more individualized, open-ended, manner and achieve a deeper scientific literacy.

There is a distinction between immersive formal and informal environments outside of school. A formal environment maintains knowledge and formal skills as its primary goal and may either act as a booster for classroom instruction (e.g. a week long SAT II preparation course on a college campus) or as an accelerator (e.g. advanced science programs, "quantum mechanics for high school students", etc...). Informal immersive environments allow for much more individual exploration, and *complement* formal education, rather than aiming to provide direct support.

What then are the components of a well designed immersive informal environment? Key aspects are attitude, immersion and program design. The attitude of the program towards science should reflect the distinction above between formal and informal programs, and aim to produce students who are more inquisitive about science, and are inquisitive in a more effective manner.

Immersion is important – the program should last for a substantial time, this can either be continuous (e.g. a week or more at a residential center) or spread over time (e.g. a semester long program meeting once a week at a museum). This allows students and instructors to get to know each other, and more importantly allow instructors to understand the questions that their students need to answer. Experience and experimentation can also take a long time as students become comfortable with materials and equipment, and discover what it is that they are interested in finding out.

A well designed environment must have a well designed underlying program. In terms of the practical science areas discussed earlier, there should be time for experimentation, exercise and investigation – though none of these need necessarily be formalized to the extent that they are in the classroom. The program design should include basic concepts and ideas that the instructor hopes their students will gain a better understanding of, and sufficient resources to allow each child to experience, experiment and investigate. A well designed program does *not* mean that a rigorous schedule or curriculum is required.

Without "pure experimentation" it is very hard for students to discover the questions that they are interested in. Even though this may look like play to the outsider, raw experience is a vital component in developing an interest in a particular phenomenon or developing a plan to use a particular piece of equipment (Woolnough & Allsop, 1985; Polman, 2000). The instructor may elect to use some exercise type instruction to directly teach some content or a formal skill, but this should only be used where essential to support the growth of understanding – exercises should not (and probably *can not*) drive a child's curiosity. Once young people have developed an interest they should be allowed time to investigate the questions that they generate. With support of their instructor the informal skill of turning knowledge into understanding can be developed.

We can not (and should not) aim to turn every child into a professional scientist. Rather, we would like every individual to be able to think and work effectively in the highly science and technology driven communities in which we live. This goal is achieved when young people learn that they can develop their own understanding in a variety of settings; this they can only learn through a diverse range of experiences including formal education (in schools and formal immersive programs), short informal experiences (visiting museums and science centers) and immersive informal programs of the kind described here.

## 5. Conclusions

The best understanding of science occurs when young people have inquiring minds that are able to convert new knowledge into genuine understanding using a variety of formal and informal skills. Science instruction ranges on a spectrum from the most informal (noticing things out in the world) to formal (learning in a classroom), all of which can have practical aspects including experiences, exercises and investigations.

Well rounded scientific thinkers have a background of science learnt in multiple venues that not only focus on factual knowledge and formal skills, but also on developing an inquisitive attitude and an inquiring mind.

Informal immersive environments, such as residential visits, summer camps and long-term hands-on programs at science centers can provide learning venues where the development of experimentation and inquiry can complement the knowledge and skills learnt in school and individuals can discover how to regularly turn knowledge into their own personal understanding by becoming well rounded, questioning, interested thinkers.

### Acknowledgements

Thanks to Neil Corbett for many useful comments on the early versions of this work.

## References

American Association for the Advancement of Science (AAAS) Project 2061, 1993, "Benchmarks for Science Literacy" (New York & Oxford: Oxford University Press)

Bybee, R. W. 2001, in "Free choice science education", Falk, J. H. ed. (New York, NY: Teachers College Press) chap. 3

Falk, J. H. 2001, in "Free choice science education", Falk, J. H. ed. (New York, NY: Teachers College Press) chap. 1

Layman, J. W. 1996, "Inquiry and Learning" (New York, NY: The College Entrance Examination Board)

Mestre, J. P., & Cocking, R. R. 2000. Journal of applied developmental psychology 21 (1):1

Mestre, J. P., & Cocking, R. R. 2002, in "Learning Science and the Science of Learning", Bybee, R. W. ed. (Arlington, VA: NSTA Press) Chapter 2

National Research Council (NRC) 1995, "National Science Education Standards" (Washington, DC: National Academy Press)

Noel-Storr, J., et al. 2002, Bulletin of the American Astronomical Society, 33:1344

Polman J. L. 2000, "Designing Project Based Science" (New York, NY: Teachers College Press)

Reardon, J. 2002, in "Science Workshop: Reading, Writing and Thinking like a scientist", Saul, W., et al., 2$^{nd}$ Ed. (Portsmouth, NH: Heinemann)

Woolnough, B. & Allsop, T., 1985, "Practical work in science" (Cambridge: Cambridge University Press)